\documentclass[10pt,twocolumn,letterpaper]{article}

\usepackage[pagenumbers]{wacv} 

\usepackage{graphicx}
\usepackage{amsmath}
\usepackage{amssymb}
\usepackage{booktabs}
\usepackage{graphicx}
\usepackage{booktabs}
\usepackage{colortbl}
\usepackage{tabularx}
\usepackage[accsupp]{axessibility} 
\usepackage[pagebackref,breaklinks,colorlinks]{hyperref}
\usepackage{orcidlink}
\usepackage{multirow}

\usepackage[pagebackref,breaklinks,colorlinks]{hyperref}

\usepackage{enumitem}

\usepackage[capitalize]{cleveref}
\crefname{section}{Sec.}{Secs.}
\Crefname{section}{Section}{Sections}
\Crefname{table}{Table}{Tables}
\crefname{table}{Tab.}{Tabs.}

\def\wacvPaperID{1} 
\def\confName{WACV}
\def\confYear{2025}

\begin{document}

\title{MSM-BD: Multimodal Social Media Bot Detection Using Heterogeneous Information}

\author{
Tingxuan Wu$^{1\dag}$, Zhaorui Ma$^{2}$, Yanjun Cui$^{3}$, Ziyi Zhou$^{3}$, and Eric Wang$^{4}$\\
$^{1}$New York University; $^{2}$George Mason University; $^{3}$Dartmouth College; \\$^{3}$Independent Researcher\\
{\tt\small tw3196@nyu.edu,} 
{\tt\small zma4@gmu.edu,}  \\
{\tt\small \{yanjun.cui.gr, ziyi.zhou.gr\}@dartmouth.edu,} \\ 
{\tt\small ericw1122@yahoo.com}
}
\maketitle

\begin{abstract}
Although social bots can be engineered for constructive applications, their potential for misuse in manipulative schemes and malware distribution cannot be overlooked. This dichotomy underscores the critical need to detect social bots on social media platforms. Advances in artificial intelligence have improved the abilities of social bots, allowing them to generate content that is almost indistinguishable from human-created content. These advancements require the development of more advanced detection techniques to accurately identify these automated entities. Given the heterogeneous information landscape on social media, spanning images, texts, and user statistical features, we propose \textbf{MSM-BD} -- A \underline{\textbf{M}}ultimodal \underline{\textbf{S}}ocial \underline{\textbf{M}}edia \underline{\textbf{B}}ot \underline{\textbf{D}}etection approach using heterogeneous information. MSM-BD incorporates specialized encoders for heterogeneous information and introduces a cross-modal fusion technology, Cross-Modal Residual Cross-Attention (CMRCA), to enhance detection accuracy. We validate the effectiveness of our model through extensive experiments using the TwiBot-22 dataset.
\end{abstract}

\newcommand\blfootnote[1]{%
  \begingroup
  \renewcommand\thefootnote{}\footnote{#1}%
  \addtocounter{footnote}{-1}%
  \endgroup
}

\blfootnote{%
\noindent\makebox[0pt][l]{$\dag$}
\parbox[t]{\dimexpr\linewidth-1em\relax}{
\hspace{1em}Project Lead\\
Selected for publication in Springer Nature Studies in Computational Intelligence
}
}


\section{Introduction}
\label{sec:1}

\begin{figure}[tb]
\centering
\resizebox{0.46\textwidth}{!}{
\includegraphics{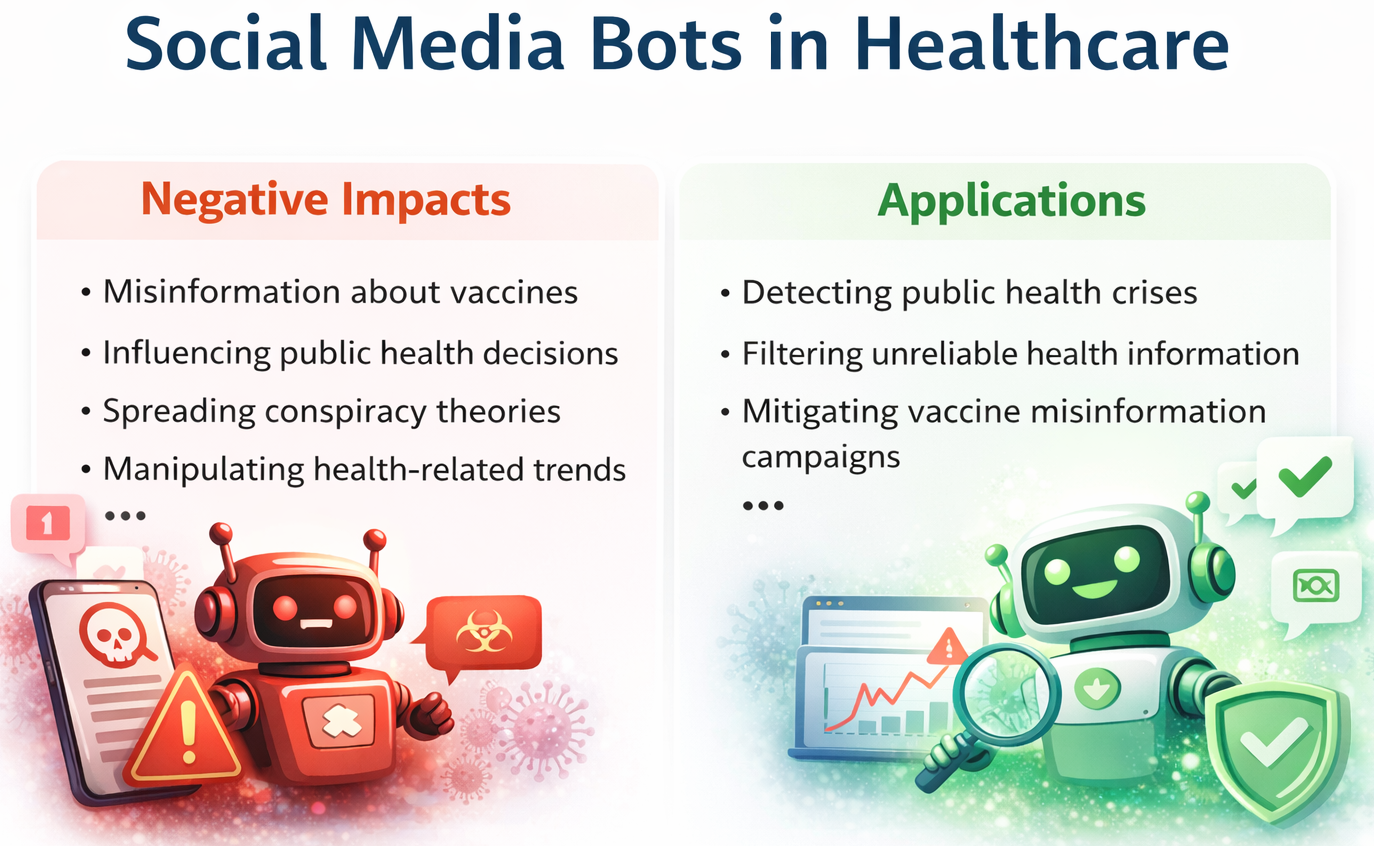}
}
\caption{Examples of negative impacts of social media bots on healthcare, such as spreading misinformation, and applications of bot detection, like public health crises detection.}
\label{fig:teaser}
\end{figure}

The rise of automated accounts, commonly known as bots, on platforms like X is a widely recognized issue. These bots, particularly malicious ones, have been associated with a variety of problems, including spreading fake news \cite{cui2020deterrent}, meddling in elections \cite{howard2016bots}, and propagating conspiracy theories \cite{ferrara2020types}. Beyond these issues, recent studies highlight the potential implications of social bots in healthcare and population medicine, as shown in Figure~\ref{fig:teaser}. These include disseminating misinformation about vaccines \cite{broniatowski2018weaponized}, influencing public health decisions \cite{vosoughi2018spread}, and manipulating health-related trends \cite{lazer2018science}. Addressing these challenges requires advanced methods to detect bots, as well as a deeper understanding of their impact in critical domains like healthcare. For instance, research has shown how detecting bots could mitigate their influence on the spread of medical misinformation \cite{ferrara2020types}, improve early detection of health-related crises \cite{shin2016high}, and support population-level interventions by filtering unreliable information \cite{carley2020social}.

Detecting social bots has attracted attention in various domains such as artificial intelligence \cite{efthimion2018supervised} and cybersecurity \cite{fazil2021deepsbd}, but it remains a formidable challenge \cite{cresci2020decade}. Similarly to a Turing test, the aim here is to determine whether a profile is managed by a human or a machine. But unlike the Turing test, which relies on direct interactions for making judgments, bot detection depends solely on the analysis of profile information. Further complicating this task, social bots exhibit a wide range of behaviors, each emulating a distinct aspect of human activity. This diversity requires detection algorithms to have the adaptability to recognize a broad spectrum of bot characteristics to ensure accurate identification.

In social media bot detection, reliably extracting effective features remains a significant challenge, particularly within the ever-evolving landscape of platform X. These bots, aiming to avoid detection, adeptly emulate real user behaviors, producing tweets that closely mirror those of authentic users. 

Consequently, traditional methods that rely on the extraction of single-modality features are inadequate to identify the nuanced characteristics that differentiate social bots. Recent advances in multimodal learning have shown that more effective modeling often depends on fine-grained cross-modal interaction, adaptive fusion, and mechanisms that separate perception from higher-level reasoning \cite{diao2025protovqa,zhang2025knowing,diao2025learning,diao2026addressing}. Despite efforts to improve detection models by incorporating heterogeneous information, accuracy remains suboptimal \cite{feng2022twibot}. Consequently, one crucial question has arisen: \textbf{Is it feasible to develop a multimodal bot detection method that effectively utilizes heterogeneous information from social media platforms for more accurate identification?}

To overcome the challenges above, we propose \textbf{MSM-BD} -- A \underline{\textbf{M}}ultimodal \underline{\textbf{S}}ocial \underline{\textbf{M}}edia \underline{\textbf{B}}ot \underline{\textbf{D}}etection approach using heterogeneous information.
In summary, our contribution is threefold:
\begin{enumerate}[leftmargin=*]
    \item {
    \textbf{End-to-end Multimodal Social Media Bot Detection.} 
    We propose MSM-BD, a one-stage end-to-end multimodal bot detection method that leverages modality-specialized feature encoders and our proposed cross-modal fusion module to effectively utilize heterogeneous information from social media.
    }
    \item {
    \textbf{Effective Cross-modal Fusion.} 
We introduce a Cross-Modal Residual Cross-Attention (CMRCA) module to effectively handle the heterogeneous information of diverse data types in various application contexts. This module significantly improves our model's capability to detect social bots by optimizing the fusion of multimodal embeddings. 
    }
    \item {
    \textbf{State-of-the-Art Performance.}   We conduct comprehensive experiments on the TwiBot-22 dataset \cite{feng2022twibot} to validate the effectiveness of MSM-BD. Our results demonstrate that our proposed model consistently detects social media bots under various conditions.
    }
\end{enumerate}

\begin{figure*}[htbp]
  \centering
  \includegraphics[width=1\textwidth]{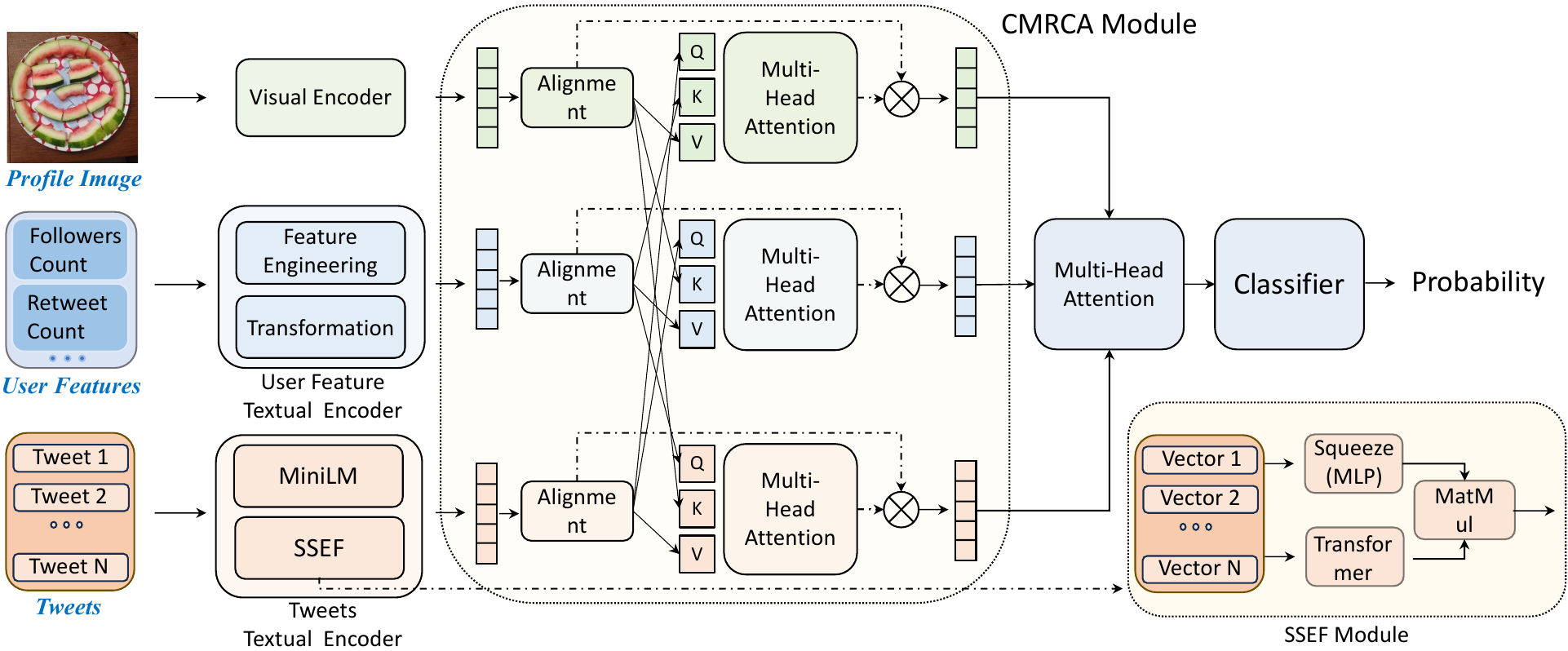}
  \caption{
  \textbf{The structure of MSM-BD pipeline.}
MSM-BD effectively utilizes profile images, user features, and tweets, processing these inputs through specialized encoders and employing CMRCA module to fuse extracted features for accurate bot detection.
  }
  \label{pipeline}
\end{figure*}

\section{Related Work}
\label{relatedwork}
Research has investigated various dimensions of social bots, including the identification of automated content \cite{almerekhi2015detecting}, detection of bot-controlled profiles \cite{2020Detection}, and reevaluation of existing detection systems \cite{2021Bot}. Studies have also explored different platforms such as Reddit \cite{hurtado2019bot} and Facebook \cite{2012Social}, which reflect the diverse environments in which bots operate. In the context of our work, we particularly emphasize the exploration of social bot detection techniques for X, as they directly align with our research goals and scope. This targeted focus allows us to delve into the intricate aspects of bot detection, providing a more granular analysis and contributing to the advancement of the field.

\subsection{Crowdsourcing Based Methods}
Crowdsourcing has become a useful approach to identify social bots \cite{2012Social}. Cresci \textit{et al.} \cite{cresci2017paradigm} conduct an in-depth analysis of crowdworkers' abilities to differentiate social bots, showcasing their ability to distinguish between typical spam bots and real human accounts. However, this study also identifies significant challenges faced by crowdworkers to reliably detect sophisticated social spam bots.

\subsection{Social Network Based Methods}
Prior research has focused on the behavioral patterns of bots,  particularly in the context of their social network expansion. Boshmaf \textit{et al.} \cite{2013Design} reveal that bots tend to accept connections indiscriminately. Golbeck \textit{et al.} \cite{2019Benford} identify deviations from Benford's law in bot friendships, thereby proposing a graph-based detection method. Niculescu-Mizil \textit{et al.} \cite{lee2011seven} use honeypot accounts to identify content polluters, tagging approximately 36,000 accounts and noting their propensity to disseminate suspicious links, including phishing sites.

\subsection{Machine Learning Based Methods}
In the domain of machine learning, methods for detecting X bots are broadly classified into feature-based, text-based, and graph-based approaches \cite{feng2022twibot}. Efforts to improve detection accuracy have involved combining tweets and user features \cite{cai2017detecting}, developing unsupervised user representations \cite{feng2021satar}, and addressing multilingual content issues \cite{knauth2019language}. However, as bots continue to evolve and better mimic authentic user content, text-based detection methods face new challenges \cite{cresci2020detecting}. In response, recent studies have begun to integrate graph-based methods \cite{guo2021social} with textual analysis and developed innovative graph neural network (GNN) models to better exploit the complex nature of the X network \cite{feng2022heterogeneity}.

\section{MSM-BD Pipeline}
\label{sec:2}
The overview of MSM-BD is shown in Figure \ref{pipeline}, which outputs the bot probability ($p$) based on the heterogeneous features: profile image features from the visual modality ($F^V_v$), user features ($F^T_u$) and tweet text features ($F^T_t$) from textual modality. These cross-modal features are then fused through a Cross-Modal Residual Cross Attention (CMRCA) module, followed by a classifier to obtain the probability. The overall pipeline is expressed as below:
\begin{equation}
    p = Classifier(CMRCA(F^V_v, F^T_u, F^T_t)),
\end{equation}
where $F^{\{V, T\}}$ indicates features from visual/textual modality, $F_{\{v, u, t\}}$ represents the domain of visual profile, user features, and tweets, respectively. $p$ represents the probability that an account is a bot.
 This probability is a critical indicator in the bot detection process, facilitating the identification and mitigation of automated bot behavior on social media platforms.

\subsection{Multimodal Encoders}
\subsubsection{Visual Encoder}
The Visual Encoder (VE) used in MSM-BD is leveraged from a ResNet-18 model \cite{he2016deep} pretrained on ImageNet \cite{russakovsky2015imagenet}. 
The pretraining process on ImageNet provides the model with a more comprehensive understanding of various visual concepts than training it from scratch, allowing it to capture meaningful features from user profile images. Therefore, given $I^V_v$ as the original profile image input, the extracted visual features $F^V_v$ enable the model to adequately learn the difference between bot accounts and real ones, which facilitates subsequent feature fusion and bot detection, as expressed below:
\begin{equation}
    F^V_v = ResNet(I^V_v).
\end{equation}

\subsubsection{User Feature Textual Encoder}
The User Feature Textual Encoder includes feature vectorization and transformation. 
Inspired by Kantepe \cite{kantepe2017preprocessing}, we use feature engineering to vectorize and extract statistical features from the original user information $I^T_u$ from various aspects, including the relevant count, status and properties. The information involved is shown in Table \ref{tab:userfeature}.

\begin{table*}[htbp]
\centering
\resizebox{1.0\linewidth}{!}{
\begin{tabular}{@{}c|c@{}}
\toprule
\multicolumn{1}{l|}{} & \textbf{User-Based Information ($I^T_u$)}                                                                                                         \\ \midrule
\textbf{Count}        & account age, tweet count, average tweets per day, followers and friends count, hashtag and mention counts                      \\
\textbf{Status}       & verification status, account protection status                                                                                 \\
\textbf{Properties}   & screen name properties, presence of descriptions, profile image attributes, location information, URL presence in descriptions \\ \bottomrule
\end{tabular}
}
\caption{User information involved in the User Feature Textual Encoder.}
\label{tab:userfeature}
\end{table*}

The encoded features are then transformed through a linear transformation followed by a non-linear GELU \cite{hendrycks2016gaussian} activation function, as expressed below:
\begin{equation}
\begin{split}
F^T_u = GELU(LT(Vectorize(I^T_u))), \\
\quad GELU(x) = x\cdot\Phi(x),
    \end{split}
\end{equation}
where $Vectorize(\cdot)$ indicates the feature engineering, $LT(\cdot)$ refers to linear transformation, and $GELU(\cdot)$ represents the non-linear activation ($\Phi(\cdot)$ is the cumulative distribution function of the standard normal distribution \cite{hendrycks2016gaussian}).  $F^T_u$ is further utilized in the CMRCA module for cross-modal feature fusion.
By incorporating this module, MSM-BD can effectively leverage user statistical features from multiple levels, improving its accuracy in social media bot detection.

\subsubsection{Tweets Textual Encoder}

MSM-BD considers the tweets information for bot detection through the proposed Tweets Textual Encoder. 
Specifically, given $I^T_t = \{I^T_{t_1}, I^T_{t_2}, ..., I^T_{t_N}\}$ as the $N$ recent tweets of the user, the tweets are first input into a pretrained language model, MiniLM \cite{wang2020minilm}, for linguistic feature extraction, as shown in Equation \ref{eqa:minilm}. As a powerful language model pretrained on an extensive corpus, MiniLM effectively captures the semantic information embedded within the text while preserving its lightweight structure. Therefore, we can obtain high-quality textual embeddings that accurately reflect the content and context of the tweets, thereby enhancing the discriminative power of our model:
\begin{equation}
\label{eqa:minilm}
    F^{T}_{l_i} = MiniLM(I^T_{t_i}), i \in [1, N],
\end{equation}
where $F^{T}_{l_i}$ indicates the linguistic features generated by MiniLM for the $i^{th}$ tweet of the user. 
To integrate these $N$ features for further fusion, a Sentence SE Fusion (SSEF) module is proposed and involved, as shown in Figure \ref{pipeline}.
Inspired by the SE Net \cite{hu2018squeeze} structure and its relevant applications in language feature extraction \cite{guo2023nuaa}, we utilize a fully connected network (MLP) to compress the extracted features of $N$ tweets ($F^{T}_{l_i}$, $i \in [1,N]$) to distill the essential information while reducing noise and redundant details. 
Meanwhile, an attention-based Transformer Encoder \cite{vaswani2017attention} is applied to enhance the embedding of each tweet by capturing the intricate relationships and dependencies within the tweet sequence. The Transformer Encoder's attention mechanism allows our model to selectively attend to relevant parts of the tweet sequence, amplifying the representation of important information while suppressing irrelevant or misleading signals. Finally, the unified embedding of the tweet sequence ($F^{T}_{t}$) is obtained through matrix multiplication which provides a comprehensive representation of the text modality. The overall process is expressed below:
\begin{equation}
\begin{split}
    F^{T}_{t} &= MLP(F^{T}_{l_1}, F^{T}_{l_2}, \ldots, F^{T}_{l_N}) \\
    &\quad \times AttnEncode(F^{T}_{l_1}, F^{T}_{l_2}, \ldots, F^{T}_{l_N}),
\end{split}
\end{equation}
where $F^{T}_{l_i}$ is the embedding generated by MiniLM, $AttnEncode(\cdot)$ represents the attention-based Transformer Encoder.

By incorporating the SSEF module, our model effectively integrates contextual information and captures fine-grained dependencies among tweets. This holistic representation enables the discrimination between bot-generated and genuine user-generated tweets, thereby enhancing the overall performance of our social media bot detection system. The combination of MiniLM-based tweet encoding and the SSEF module provides a robust and versatile framework for analyzing textual modalities in social media, with implications for detecting various forms of suspicious or malicious activity.

\subsection{Cross-Modal Residual Cross-Attention}
\label{sec:cmrca}
The heterogeneous features of multimodality ($F^V_v$, $F^T_u$, $F^T_t$) are fused through the novel Cross-Modal Residual Cross-Attention (CMRCA) module. 
Inspired by previous multimodal fusion studies \cite{tan2019lxmert,diao2024learning,diao2025temporal,li2023blip,diao2023av,diao2023ft2tf}, our approach differentiates itself by enabling an explicit definition of the roles played by different modal embeddings during the fusion process.
Specifically, we first involve a separate fully connected network to align the features of each modality, which guarantees the compatible format of the embeddings for subsequent operations. 

\begin{table*}[htbp]
\centering
\small
\setlength{\tabcolsep}{35pt} 
\renewcommand{\arraystretch}{1.2} 
\begin{tabular}{lccc}
\toprule
\textbf{Method} & \textbf{Accuracy $\uparrow$} & \textbf{\(F_1\) score $\uparrow$} & \textbf{Type} \\
\midrule
Abreu \textit{et al.} \cite{abreu2020twitter} & 0.7066 & 0.5344 & F \\
Efthimion \textit{et al.} \cite{efthimion2018supervised} & 0.7408 & 0.2758 & FT \\
Kantepe \textit{et al.} \cite{kantepe2017preprocessing} & 0.7640 & 0.5870 & FT \\
RoBERTa \cite{liu2019roberta} & 0.7207 & 0.2053 & FT \\
SGBot \cite{yang2020scalable} & 0.7508 & 0.3659 & FTG \\
RGT \cite{feng2022heterogeneity} & 0.7647 & 0.4294 & FTG \\
BotRGCN \cite{feng2021botrgcn} & 0.7966 & 0.5750 & FTG \\
\midrule
\textbf{MSM-BD (Ours)} & \textbf{0.8002} & \textbf{0.6105} & \textbf{FTI} \\
\bottomrule
\end{tabular}
\caption{Comparison with State-of-the-Art on TwiBot-22.}
\label{tab:comparison_sota_twibot22}
\end{table*}

The aligned embeddings are fed into three distinct multi-head attention mechanisms, where the Value (V) corresponds to the embedding of each modality. The roles of Key (K) and Query (Q) are assigned based on domain-specific considerations. For instance, in the user feature domain ($F_u$), user features ($F_u^T$) serve as V, tweet embeddings ($F_t^T$) as Q, and profile image embeddings ($F_v^V$) as K, leveraging their relevance to bot detection. Similar assignments apply for cross-attention in the visual ($F_v$) and tweet textual ($F_t$) domains. The single-head cross-attention can be expressed as:
\begin{equation}
\label{singlehead}
\begin{split}
        F^{f_s}_u = Softmax(\frac{F^T_t \cdot F^V_v}{\sqrt{d}})F^T_u, \\ 
        F^{f_s}_v = Softmax(\frac{F^T_t \cdot F^T_u}{\sqrt{d}})F^V_v, \\ 
        F^{f_s}_t = Softmax(\frac{F^T_u \cdot F^V_v}{\sqrt{d}})F^T_t,
    \end{split}
\end{equation}
where $d$ denotes the dimension of features after alignment, and $F^{f_s}_{\{u,v,t\}}$ indicates the single-head fusion modality of each domain (image profile, user features, and tweets). 
Specifically, to enhance the representational capacity, CMRCA employs multi-head attention. Therefore, the single-head attention is further concatenated across multiple attention heads as:
\begin{equation}
\label{eq:multihead}
\begin{split}
    F^{f_m}_u &= Concat(F^{f_s}_{u_1}, F^{f_s}_{u_2}, ..., F^{f_s}_{u_M})W_u, \\ 
    F^{f_m}_v &= Concat(F^{f_s}_{v_1}, F^{f_s}_{v_2}, ..., F^{f_s}_{v_M})W_v,\\
    F^{f_m}_t &= Concat(F^{f_s}_{t_1}, F^{f_s}_{t_2}, ..., F^{f_s}_{t_M})W_t,
    \end{split}
\end{equation}
where $F^{f_m}_{\{u,v,t\}}$ denotes the $M$-head multi-head attention, $W_{\{u,v,t\}}$ is the learnable weight matrices.
By explicitly defining the roles of different modal embeddings, CMRCA allows the model to leverage the unique strengths of each modality and facilitate effective information exchange and integration.

To mitigate the risk of overfitting and preserve the original features of each modality, we employ residual connection after the multi-head attention mechanism through the concatenation between the original feature and the one after attention:
\begin{equation}
\begin{split}
    F^{f'_m}_u = Concat(F^{f_m}_u, F^T_u),\\
    F^{f'_m}_v = Concat(F^{f_m}_v, F^V_v), \\
    F^{f'_m}_t = Concat(F^{f_m}_t, F^T_t).
    \end{split}
\end{equation}

The features are then input into a multi-head attention for integration followed by a classifier to generate the final predictions:
\begin{equation}
    p = Classifier(MultiHeadAttn(F^{f'_m}_u, F^{f'_m}_v, F^{f'_m}_t)),
\end{equation}
where the $MultiHeadAttn(\cdot)$ operation is similar to that in Equation \ref{eq:multihead}, and the classifier is linear transformation mapping from $d$-dimension vector into numerical predicted value. 

\begin{figure*}[th]
  \centering
  \resizebox{1\textwidth}{!}{%
    \includegraphics{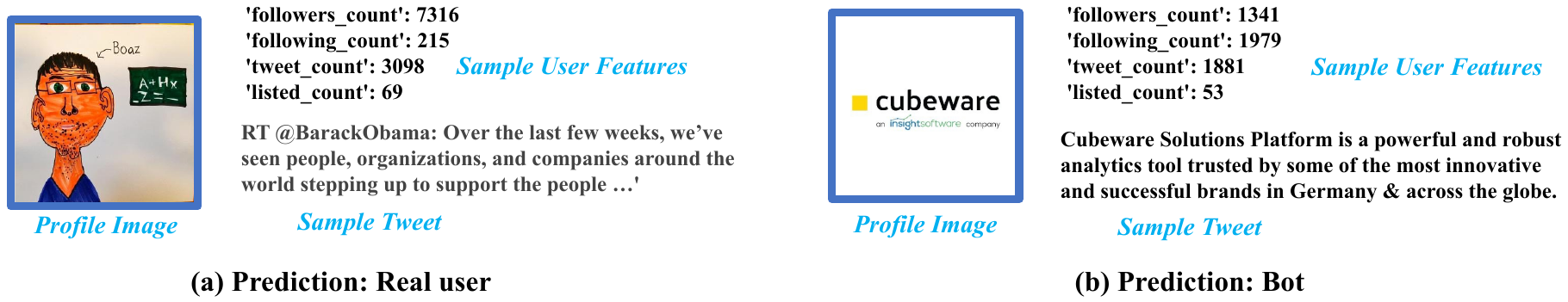}
  }
  \caption{\textbf{Demonstration of bot detection on TwiBot-22 \cite{feng2022twibot} dataset.}
  MSM-BD classifies real users and bots correctly under complicated scenarios.}
  \label{fig:sample}
\end{figure*}

\subsection{Dataset}
We use the widely used TwiBot-22 dataset \cite{feng2022twibot} to investigate the challenge of localizing X bots. TwiBot-22 consists of a diverse collection of 1,000,000 users, covering 860,057 human accounts and 139,943 bot accounts. 
The large-scale and diverse feature information (\textit{e.g.} users, tweets, hashtags, \textit{etc.}) collected by this dataset makes it ideal for bot detection task. To maintain consistency and allow meaningful comparisons with previous studies, we adhere to the original train, validation, and test splits of the dataset \cite{feng2022twibot}. This ensures fairness and allows for a comprehensive evaluation of our findings with existing research.

We expand the scope of the data modals with new categories based on the TwiBot-22 dataset \cite{feng2022twibot}. Each method is assigned to one or more categories: ``F'' (user metadata and feature engineering), ``T'' (tweet and user description text analysis), ``G'' (network structure analysis of X), and ``I'' (profile image analysis), based on specific criteria. To be specific, methods that use user metadata and feature engineering are categorized as ``F'', while those that analyze tweets and user description texts are under ``T''. Additionally, methods focusing on network structure analysis of X are categorized as ``G'', and those that leverage profile images fall under ``I''.

\subsection{Comparison with State-of-the-Art}

Table~\ref{tab:comparison_sota_twibot22} provides a comprehensive comparison of the performance of different methods on the TwiBot-22 test set. We employ the Accuracy and \(F_1\) score metrics to evaluate the effectiveness of the models in social media bot detection. Notably, MSM-BD incorporates features from multiple modalities, including user metadata (``F''), text (``T''), and images (``I''), resulting in its comprehensive categorization of ``FTI''. This multimodal approach leverages the strengths of different modalities, enhancing the model's capability to detect social media bots.

MSM-BD demonstrates superior performance with an accuracy of 0.8002 and an \(F_1\) score of 0.6105, outperforming all other methods on the TwiBot-22 dataset. MSM-BD achieves a 13.26\% increase in accuracy and a 14.26\% increase in \(F_1\) score over the best ``F'' method by Abreu \textit{et al.} \cite{abreu2020twitter}. Compared to the best ``FT'' method by Kantepe \textit{et al.} \cite{kantepe2017preprocessing}, MSM-BD shows improvements of 4.74\% in accuracy and 4.00\% in the  \(F_1\) score, illustrating its enhanced predictive precision. Additionally, MSM-BD surpasses the leading ``FTG'' method, BotRGCN \cite{feng2021botrgcn}, with a 0.45\% increase in accuracy and a 6.17\% increase in the \(F_1\) score. Both metrics highlight MSM-BD's robust capability to accurately classify bots in social media, showcasing its effectiveness across different types of method with substantial improvements, particularly in the \(F_1\) score.

Figure \ref{fig:sample} shows two samples from the TwiBot-22 dataset. In these complex scenarios, where both samples have more than 1000 followers and a high volume of tweets, our model can accurately differentiate between real users and bots.

MSM-BD distinguishes itself by incorporating image-based features, providing an additional layer of discriminatory power. This integration of diverse data types allows our model to outperform baseline models, highlighting the critical importance of multimodal analysis in detecting social media bots. The strength of MSM-BD lies in its ability to capture heterogeneous information from various modalities and effectively fuse them to enhance detection performance.

\subsection{Ablation Study}
We conduct an ablation study of MSM-BD on the TwiBot-22 \cite{feng2022twibot} dataset to investigate the impact of our proposed CMRCA module. As discussed in \cite{diao2023ft2tf}, we assume the cross-attention module will improve model's representation learning capability. The results are shown in Table \ref{tab:ablation}.

\begin{table}[htbp]
\centering
\setlength{\tabcolsep}{12pt} 
\renewcommand{\arraystretch}{1} 
\resizebox{1\linewidth}{!}{ 
\begin{tabular}{lcc}
\toprule
\textbf{Method} & \textbf{Accuracy $\uparrow$} & \textbf{\(F_1\) score $\uparrow$} \\
\midrule
w/o CMRCA module & 0.7204 & 0.4516  \\
\textbf{MSM-BD (Ours)} & \textbf{0.8002} & \textbf{0.6105} \\
\bottomrule
\end{tabular}
}
\caption{Ablation study on the CMRCA module of MSM-BD on TwiBot-22. ``w/o CMRCA module'' indicates the model without the CMRCA module, using concatenation followed by a fully connected layer instead.}
\label{tab:ablation}
\end{table}

Excluding the CMRCA module, the model combines outputs from the individual modality-specialized encoders (Visual Encoder, User Feature Textual Encoder, and Tweets Textual Encoder) using concatenation followed by a fully connected layer, instead of the attention-based fusion provided by CMRCA. The result of the ablation study demonstrates the significant contribution of the CMRCA module to the overall performance of MSM-BD.

\section{Conclusion}
We present MSM-BD, a multimodal social media bot detection model that effectively utilizes heterogeneous information from images, text, and user statistics. MSM-BD overcomes the limitations of traditional methods that rely on single-modality features, providing a more comprehensive and accurate detection capability. Our experiments on the TwiBot-22 dataset demonstrate that MSM-BD significantly improves bot detection performance. This work offers a solution to enhance the accuracy and reliability of social media bot detection, inspiring further research to integrate diverse data sources for more effective solutions.

{\small
\bibliographystyle{ieee_fullname}
\bibliography{egbib}
}

\end{document}